# Grid Technologies

**Univ.Assist. Delia Sabina Stîngă, Ph.D.Candidate**
**Faculty of Computers and Applied Computer Science**
**"Tibiscus" University of Timişoara, România**

**ABSTRACT.** This paper contains the most important aspects of computing grids. Grid computing allows high performance distributed systems to act as a single computer. An overview of grids structure and techniques is given in order to understand the way grids work.

## 1. Introduction

A computing grid is a software and hardware infrastructure that allows reliable, consistent, wide spread and cheap access at the high-end capacities of the computers.

Infrastructure: because a computing grid is focused, above all, on a large scale sharing of resources and also on computing cycles, data or people. This sharing involves an important hardware infrastructure in order to obtain the necessary interconnections and a software infrastructure to survey and control all the results.

Wide access allows us to rely on always available services, in any environment we wish to act. But this does not mean that the resources are everywhere or that they are universal available.

The first step in performing a computing that involves shared resources is the *authentication process*, designed to establish the user's identity. A further authentication process establishes the user's rights to create entities called *processes*. A process contains one or more threads created for competition or parallelism and executed in the frame of a *shared addressing space*. A process can communicate with other processes via a variety of abstractions, including shared memory (with semaphores or locks), pipes and protocols like TCP/IP. A user (or a process activated by a user) can *control* the activities from another process – for example, can suspend, restart or terminate its execution. This control is accomplished via the method of asynchronous delivered signals. A process acts like on behalf of the user to *collect resources* by executing instructions, occupying memory, reading or writing disks, sending or receiving messages, etc. The ability of a process to collect resources is restricted by the involved authorization mechanisms, which implements a system resources allocation politics, by taking in consideration the user's identity, the consuming of the previous resource,… The *planning mechanisms* from the involved system deal with the competitive resource demands and can also support the user's demands for performance guarantees. The *accounting mechanisms* track the resource allocation and consuming. The *payment mechanisms* can be offered to translate the resource consuming in several common currency. The system will also provide *protection mechanisms* in order to guarantee the fact that the user's computing will not interfere with others. Other services provide abstractions for secondary storage: *virtual memory*, *file systems* and *databases*.

A major characteristic for the techniques used for the implementation of the grid services is the scale. Adopting the scale as the major comparison dimension, there are four system types, each of increasing scale and complexity. These are:

1. Final systems (computers, storage systems, sensors and other devices) - These are characterized by a relatively small scale and a high degree of homogeneity and heterogeneity.
2. Clusters – The network of workstations: o collection of computers connected via a high speed local network and designed to be used as an integrated computing resource or as data processing. Like the final system, the cluster is a homogenous entity, its systems being different in configuration and not in basic architecture. It is controlled by a single administrative entity that has total control on every final system. Clusters are characterized by increased physical scale and reduced integration. These characteristics make the implementation of some services difficult and there is the need of new services that are not requested in a single final system.
3. Intranet – The Intranet is a grid that includes a possible big number of resources, although they belong to a single organization. Like a cluster, the intranet can involve central administrative control, therefore a high degree of coordination between the resources. It is characterized by heterogeneity (final systems and the networks used in an intranet are usually of different types and capabilities), by separate administration and by the lack of global knowledge. It is not possible for a person or a computation to have an accurate global knowledge of the system's structure or state.

4. Internet – This is the most challenging of the system classes in which networking computing can be performed. Internet is an inter-network system that extends over multiple organizations. It tends to get wide and heterogeneous. The Internet is characterized by the lack of centralized control. There is no authority that can impose operational politics or that can ensure the quality of the resources, so there will be a great variety in what concerns the politics and quality. Another characteristic: the geographical distribution. Typically, the Internet connects resources that are wide area geographical distributed. If a grid lies over international borders, the export controls can restrain the technologies that can be used for security, etc.

## 2. Grid Middleware

The collection of APIs, protocols and software that allows to create and use a distributed system represents a grid middleware. It stands on a lower level than the user's application, but n a higher level than the network transport protocols. There is a great variety of middleware packages, a lot of projects are developing middleware that allow creating some production grids. The most important project is the Globus project. Some other important projects are Legion and Condor, Akenti, NetSolve, Ninf.

## 3. Globus Toolkit

A grid software infrastructure's approach is represented by the Globus project. Globus develops an integrated set of grid basic services, the set being called Globus toolkit. Globus is different than the other architectures by three aspects: by its services model that can allow the applications to use the grid services without adopting a particularly programming model; by its specialized mechanism that can coexist with (sometimes replace) the mechanism provided by the commodity computing; and by the support offered to a application high demands process approach based on information.

    The toolkit has a set of components that implements the basic services for security, resource allocation, resource management, communication, etc.

    The Globus services are listed below:

| Service | Name | Description |
|---|---|---|
| Resource administration | GRAM | Resource allocation and process management |
| Communication | Nexus | Unicast and multicast communication services |
| Information | MDS | Distributed access to information structure and state |
| Security | GSI | Authentication and security services |
| Health and status | HBM | Surveying the systems' components health and status |
| Data remote access | GASS | Data remote access via sequential and parallel interfaces |
| Exe management | GEM | Exe building, keeping in cache memory and localization |

## 4. Resource Allocation

The high level global services are build over an essential set of local services. The Globus Resource Allocation Administrator (GRAM) provides the local components for the resource administration. The GRAM is responsible for the set of resources that respect the same allocation politics. For example, a single administrator can provide access to a parallel computer, a cluster or a set of machines nodes. Although, a grid build with the help of Globus usually has more GRAMs, each of them being responsible for a particular "local" set of resources.

The GRAM provides a standard network interface for the local resource administration systems. This is the reason why grid tools and applications can express their resource allocation and process administration demands in standard API terms. The resource demands are expressed in the terms of an extensible resource specification language (Resource Specification Language - RSL) that plays a critical part in the definition of global services.

## 5. Communication

The Globus toolkit communication services are provided by the Nexus communication libraries. Nexus defines a low level communication API that is used for dealing with a great variety of higher level programming models, including message transmitting, remote procedure call, remote I/O and maintaining the shared state in the collaborative environments.

Communication in Nexus is defined in the terms of two basic abstractions. A communication link is has a communication start point linked to a communication end pint; a communication process is initialized applying remote service request (RSR) to a start point. This asynchronous and remote procedure call transfers data from the start point to the end point(s) associated and then integrates the data in the processes that contains the end point(s) calling a function in the process (es). More than a single start point can be linked to an end point and vice versa, considering the communication complex structure.

## 6. Globusrun

Globusrun is used to present jobs to the grid resources. Job's startup is made using the GRAM or DUROC Globus services. The GASS service can be also used to provide access to remote files and to redirect standard output strings. To start a job, globusrun can be used to list the jobs started before, job status requestes, to parse the RSL requests strings and to execute authentication tests for the GRAM gatekeepers. Globusrun implies the existence of a valid proxy for all the esential operations.
globusrun [-help] [-usage] [-version]
globusrun options [-r resources] [--] rsl-string
globusrun options [-r resources] –f rsl-file
globusrun –l[ist]
globusrun –status jobId
globusrun –k[ill] jobId

## 7. Java Commodity Grid Kit

Commodity grid project creates as a community's effort what it is called the Commodity Grid Toolkit (CoG Kits) that defines the mapping and interfaces between the grid services and particular commodity frameworks. Java Cog Kit is general enough to be used to design a variety of grid applications with different user request. Java CoG Kit integrates Java and the grid components in a single toolkit, as a set of services and components. The Java CoG Kit purpose is to allow grid developers to use as many of Globus Toolkit's functionalities as they want and to have access to many additional libraries and frameworks developed by the Java community. Java CoG Kit offers the user access to these grid services:
- an information service compatible with Globus Toolkit Metacomputing Directory Service (MDS), implemented using JNDI
- a security infrastructure compatible with Globus Toolkit Grid Security Infrastructure (GSI), implemented using the security library IAIK
- a data transfer compatible with a subset of Globus Toolkit gridFTP and/or GSIFTP
- resource management and jobs presentation Globus Resource Access Manager (GRAM)
- a certificate magazine based on MyProxy server

## 8. Grid Services

Java Cog contains a set of components that demonstrates the kit's utility as a base for the graphic grid applications and offers a convenient interface of the low level grid client applications.

Configuration applications - offers the user a way to configure the Java Cog Kit.

Grid Proxy Init - offers a visual interface to create the Globus proxy.

MyProxy - is used to run the proxys on MyProxy servers.

MDS components – they are simple examples that can be used to develop more sophisticated components.